\documentclass[aip,jcp,amsmath,amssymb,reprint,numeric]{revtex4-1}
\usepackage{graphicx}
\usepackage{subfigure}
\usepackage{color}
\usepackage{amsmath}
\usepackage{bm}

\begin{document}

\title {Validity of the Rosenfeld relationship: A comparative study of the network forming NTW model and other simple liquids.}
\author{Atreyee Banerjee}
\thanks{ A. Banerjee and M. K. Nandi contributed equally to this work.}
\affiliation{\textit{Polymer Science and Engineering Division, CSIR-National Chemical Laboratory, Pune-411008, India}}

\author{Manoj Kumar Nandi}

\affiliation{\textit{Polymer Science and Engineering Division, CSIR-National Chemical Laboratory, Pune-411008, India}}

\author{Sarika Maitra Bhattacharyya}
\email{mb.sarika@ncl.res.in}
\affiliation{\textit{Polymer Science and Engineering Division, CSIR-National Chemical Laboratory, Pune-411008, India}}

\date{\today}

\begin{abstract}

\textbf{Abstract.} In this paper we explore the validity of the  Rosenfeld and the Dzugutov relation for the Lennard-Jones (LJ) system, its repulsive counterpart, the WCA
system and a network forming liquid, the NTW model.
We find that for all the systems both the relations are valid at high temperature regime with an universal exponent close to 0.8.
Similar to that observed for the simple liquids, the LJ and the WCA systems show a breakdown of the scaling laws at the low temperature regime.
However for the NTW model, which is a simple liquid, these scaling laws are valid even at lower temperature regime similar to that found for ionic melts.
Thus we find that the NTW model has  mixed characteristics of  simple liquids and  ionic melts.
Our study further reveals a quantitative relationship
between the Rosenfeld and the Arrhenius relations. 
For strong liquids, the validity of the Rosenfeld relation in the low temperature regime is connected to it following the Arrhenius behaviour in that regime.
Finally we explore the role of pair entropy and residual multiparticle entropy in the dynamics
 as a function of fragility of the systems.

\end{abstract}
\keywords{ Rosenfeld scaling, Arrhenius law, excess entropy, pair entropy}
\maketitle

\section{Introduction}

A semi-quantitative relation between dynamical properties like diffusivity or relaxation times and the thermodynamics  has been proposed
by Rosenfeld \cite{Rosenfeld,Rosenfeld-iop} and recently it has been extensively studied for different systems\cite{Goel_charu,murari_charu_2010_jcp,Ruchi_charu_2006,
manish_charu,Murari_charu_jcp,agarwal2009relationship,dhabal2016comparison}. The relationship suggests that the fluid should follow $X^*=C \exp [-K S_{ex}]$,
where $X^*$ is the dimensionless dynamical quantity and $S_{ex}$ is excess entropy which is the difference
between the total thermodynamic entropy ($S_{tot}$) and the corresponding ideal gas entropy ($S_{id}$) at
the same temperature ($T$) and density ($\rho$). C and K are the constants in the Rosenfeld relation.
In a liquid due to the structural correlation the total entropy is less than the corresponding ideal gas entropy. Thus, the excess entropy has a negative
value. It can also be expanded in 
an infinite series, $S_{ex}= S_2+S_3+....S_n$, where $S_n$ is the entropic contribution due to n-particle spatial correlation\cite{Kirkwood}. 
For a binary system, the entropic contribution due to the pair correlation ($S_2$) can be written in terms of 
the partial radial distribution functions,

\begin{equation}
\frac{S_{2}}{k_{B}}=-\frac{\rho}{2} \sum_{\alpha,\beta}x_{\alpha} x_{\beta} \int_0^{\infty} \{g_{{\alpha}{ \beta}}(r) \ln g_{{\alpha} {\beta}}(r)- [g_{{\alpha}{\beta}}(r)-1]\} d {\bf r},
\label{s2_final}
\end{equation}
\noindent  where $ g_{{\alpha}{ \beta}}(r)$ is the  pair correlation between particles of type $\alpha$ and $\beta$, $\rho$ is the density of the system,
 $x_{\alpha}$ is the mole fraction of component $\alpha$ in the mixture, and $k_B$ is the Boltzmann constant.
Thus the excess entropy can be written as a contribution from the pair, $S_2$ and higher order terms, $\Delta S$. The $\Delta S$ contains all the contribution
beyond two particles\cite{giaquinta-1,giaquinta-2,giaquinta-3,giaquinta-4} and  is written as,
\begin{equation}
 \Delta S= S_{ex}-S_2=\sum_{n=3}^{ \infty}S_{n},
\end{equation}
\noindent where, $S_n$ is the entropic contribution due to n-particle spatial correlation\cite{Kirkwood}.

In liquid state theory, the structure of a liquid is primarily described by radial distribution function and hence its 
entropic contribution plays a crucial role in predicting the dynamical quantities.  For simple liquids the pair entropy $S_{2}$ provides $80\%-90\%$ contribution to
 the excess entropy \cite {Baranyai-cp, murari_charu_2010_jcp}.
  Based on this  microscopic picture,  Dzugutov proposed a relationship between pair excess entropy $S_2$ and dynamical quantity which 
is given by $X^*=C_1 \exp [-K_1 S_{2}]$\cite{Dzugutov}.  Therefore, the Dzugutov scaling law is considered as a special case of the Rosenfeld scaling.
 
 Although the Rosenfeld relation is semi-empirical in nature, Bagchi and coworkers have derived it using the connection between excess entropy and Zwanzig's rugged energy landscape model of diffusion
 \cite{zwanzig,saikat_jcp,bagchi_seki}.
Samanta {\it et al.} have found that the mode coupling diffusion coefficient can be fitted to the Rosenfeld scaling under certain approximations\cite{swapanghoshprl}. 
In a recent study we find that
mode coupling theory (MCT) vertex is related to pair excess entropy and higher order MCT calculation in the schematic limit can provide the phenomenological
Rosenfeld relation \cite{unravel}. Thus, the scaling law has been derived using different theories.

   The Rosenfeld scaling relation is found to be valid for a  wide variety of liquids including simple liquids \cite{murari_charu_2010_jcp},
water\cite{agarwal2010relationship}, ionic melts
\cite{agarwal2009relationship,Ruchi_charu_2006,manish_charu},
 model polymeric melts\cite{Goel_charu} and even for the data obtained in
different experiments \cite{abramson2007viscosity,abramson2008viscosity,abramson2009viscosity}. However
the Rosenfeld behaviour is not same for all the systems.
For most of the systems like simple liquids and water
the scaling breaks down in the low temperature regime.
For ionic liquids, the Rosenfeld relationship is found to be valid up to much lower temperature and 
the value of the Rosenfeld exponents are different from that found for simple liquids\cite{agarwal2009relationship}.
The ionic melts are known to be network forming liquids. Recently Coslovich and Pastore have proposed a new model of a simple liquid (NTW model)
which mimics the properties like the structure and the dynamics
of the network forming liquids \cite{coslovich_pastore_jpcm_ntw,szamel-ntw}. Since earlier studies have reported the Rosenfeld behaviour to be 
different for simple and network forming liquids\cite{Murari_charu_jcp,agarwal2009relationship}, 
it will be interestingly to investigate the Rosenfeld scaling of the NTW model.

 In this present work, using molecular dynamics simulations we perform a comparative study of the validity of the Rosenfeld relation for
 a few simple liquids, the LJ system, its repulsive counterpart known as WCA (Weeks-Chandler-Andersen)
system and the NTW model. The earlier study of the Rosenfeld relationship for the LJ and WCA systems have been performed in a moderate density range ($1.05-1.25$)
\cite{Murari_charu_jcp}.
Here we study a higher density range from 1.2 to 1.6 and find that
 the LJ and WCA systems   
 follow the Rosenfeld and Dzugutov scaling laws only up to moderately high temperature.  On the other hand, similar to that found for  ionic melts\cite{agarwal2009relationship} 
 the Rosenfeld relationship for the NTW model is valid upto a much lower temperature. However, the Rosenfeld exponent at high temperature appears
 to be similar to that found for simple liquids which is different from the value obtained for the ionic melts\cite{agarwal2009relationship}. Thus our result shows that
 the NTW model has mixed characteristics of simple liquids and ionic melts. We show
a connection between the validity of the Rosenfeld relation and the Arrhenius behaviour. Our study predicts that for the NTW model the validity
of the Rosenfeld relation at the low temperature is connected to it being a strong liquid. We also study the independent role of pair entropy and
residual multiparticle entropy on the dynamics as function of fragility of the system.

 The paper is organized as follows:
The simulation details are given in Sec. II. In Sec. III we describe
the methods used for evaluating the various quantities of interest and provide other necessary background. Sec. IV contains a discussion of
 presented results and Sec. V contains a brief conclusion.

\section{Simulation Details}
In this study, we perform extensive molecular dynamics simulations for three-dimensional binary mixtures in 
the canonical ensemble. The system contains $N_A$ particles of type A and $N_B$ particles of type B under periodic boundary conditions. The total number 
density is fixed at $\rho=N/V$ with the total number of particles $N = N_A + N_B$ and a system volume $V$ .
The models studied here, are the well-known models of glass-forming liquids: the binary Kob-Andersen Lennard-Jones (KALJ) liquids \cite{kob}and  the 
corresponding WCA version (KAWCA) \cite{chandler} and a network-forming (NTW) \cite{coslovich_pastore_jpcm_ntw}
liquid that mimics
$SiO_2$ with short-range spherical potentials. The molecular dynamics (MD) simulations have been carried out using the LAMMPS 
package \cite{lammps}. For all state points, three to five independent samples with run lengths $>$ 100$\tau$ ($\tau$ is the $\alpha$-
relaxation time) are analyzed.

\subsection{ KALJ and KAWCA: binary mixture of Kob Andersen Lennard-Jones particles and corresponding WCA version}

 The most frequently studied model for glass forming liquids is Kob-Andersen model which is a binary mixture (80:20) of Lennard-Jones (LJ) particles \cite{kob}.
 The interatomic pair  
potential between species $\alpha$ and $\beta$, with ${ \alpha,\beta}= A,B$, 
$U_{\alpha\beta}(r)$ is described by a shifted and truncated Lennard-Jones (LJ) potential, as given by:
\begin{equation}
 U_{\alpha\beta}(r)=
\begin{cases}
 U_{\alpha\beta}^{(LJ)}(r;\sigma_{\alpha\beta},\epsilon_{\alpha\beta})- U_{\alpha\beta}^{(LJ)}(r^{(c)}_{\alpha\beta};\sigma_{\alpha\beta},\epsilon_{\alpha\beta}),    & r\leq r^{(c)}_{\alpha\beta}\\
   0,                                                                                       & r> r^{(c)}_{\alpha\beta}
\end{cases}
\label{LJ_pot}
\end{equation}

\noindent where $U_{\alpha\beta}^{(LJ)}(r;\sigma_{\alpha\beta},\epsilon_{\alpha\beta})=4\epsilon_{\alpha\beta}[({\sigma_{\alpha\beta}}/{r})^{12}-({\sigma_{\alpha\beta}}/{r})^{6}]$ and
 $r^{(c)}_{\alpha\beta}=2.5\sigma_{\alpha\beta}$ for the LJ systems (KALJ) and $r^{(c)}_{\alpha\beta}$  is equal to the position of the minimum of $U_{\alpha\beta}^{(LJ)}$
for the WCA systems (KAWCA) \cite{chandler}. Length, temperature and
time are given in units of $\sigma_{AA}$, ${k_{B}T}/{\epsilon_{AA}}$ and $\tau = \surd({m_A\sigma_{AA}^2}/{\epsilon_{AA}})$, 
respectively.  Here we have simulated Kob Andersen Model  
with the interaction parameters  $\sigma_{AA}$ = 1.0, $\sigma_{AB}$ =0.8 ,$\sigma_{BB}$ =0.88,  $\epsilon_{AA}$ =1, $\epsilon_{AB}$ =1.5,
 $\epsilon_{BB}$ =0.5, $m_{A}$ = $m_B$=1.0. We have performed MD simulations in the canonical ensemble (NVT) using  Nos\'{e}-Hoover thermostat  with integration timestep 0.005$\tau$.
 The time
constants for  Nos\'{e}-Hoover thermostat  are taken to be 100  time steps.
The sample is kept in a cubic box with periodic boundary condition.
 System size is $N = 500$, $N_A = 400$ (N $=$ total number
of particles, $N_A$ $=$ number of particles of type A) and we have studied a broad range of density $\rho$ from 1.2 to
1.6 with different temperature ranges given in Table \ref{temp-ref}.

\subsection{NTW: tetrahedral network-forming liquids}

We study a model of network-forming liquids \cite{coslovich_pastore_jpcm_ntw} interacting via spherical short-ranged potentials.
This model is simple model and mimics $SiO_2$ glasses, in which tetrahedral networks strongly dominate the dynamics.
The interaction potentials are given as

\begin{equation}
 U_{\alpha\beta}(r)=\epsilon_{\alpha\beta}[(\frac{\sigma_{\alpha\beta}}{r})^{12}-(1-\delta_{\alpha\beta})(\frac{\sigma_{\alpha\beta}}{r})^6].
\end{equation}
Here $\delta_{\alpha\beta}$ is the Kronecker delta function. The interaction is truncated and shifted at $r=2.5\sigma_{\alpha\beta}$. The size, mass, and energy
ratios are given as  $\sigma_{AB}/\sigma_{AA}=0.49, \sigma_{BB}/\sigma_{AA}=0.85, m_B/m_A=0.57, \epsilon_{AB}/\epsilon_{AA}=24
, \epsilon_{BB}/\epsilon_{AA}=1$.

System size is $N = 500$, $N_A = 165$ (N $=$ total number
of particles, $N_A$ $=$ number of particles of type A). The number density of NTW mixtures is
$\rho = 1.655$.
These parameters are adjusted to reproduce
the radial distribution functions of the $SiO_2$ amorphous
states 
 which  corresponds to the density $\rho = 2.37g\AA{}^{-3}$
of the so-called van Beest-Kramer-van Santen (BKS) model for
the silica glass \cite{van1990force,carre2007amorphous}.

\begin{table}
 \caption{The temperature range simulated for different systems.}
\begin{center}
    \begin{tabular}{ | l | l |}
    \hline
    system (density) & Temperature Range\\ \hline
    LJ(1.2) &5.00-0.45\\ \hline
    WCA(1.2) &5.00-0.30 \\ \hline
    LJ(1.4) &5.00-1.00 \\ \hline
    WCA(1.4) &5.00-0.95 \\ \hline
    LJ(1.6) &5.00-1.89\\ \hline
    WCA(1.6) &5.00-1.89\\ \hline
 
    NTW(1.655) &1.00-0.31 \\ \hline
    \end{tabular}
\end{center}
\label{temp-ref}
\end{table}

\section{Definitions and Background}
\subsection{Relaxation time}
We have calculated the relaxation times from the decay of the
overlap function q(t), from the condition $q(t = \tau_{\alpha})/N =
1/e$.  $q(t)$ is defined as
\begin{eqnarray}
\langle q(t) \rangle \equiv \left \langle \int dr \rho(r, t_0 )\rho(r, t + t_0 )\right \rangle \nonumber\\
=\left \langle \sum_{i=1}^{N}\sum_{j=1}^{N} \delta({\bf{r}}_j(t_0)-{\bf{r}}_i(t+t_0)) \right \rangle \nonumber\\
=\left \langle \sum_{i=1}^{N} \delta({\bf{r}}_i(t_0)-{\bf{r}}_i(t+t_0)) \right \rangle \nonumber\\
+\left \langle \sum_{i}\sum_{j\neq i} \delta({\bf{r}}_i(t_0)-{\bf{r}}_j(t+t_0)) \right \rangle.
\end{eqnarray}
The overlap function is a two-point time correlation
function of local density $\rho(r, t)$. It has been used in
many recent studies of slow relaxation \cite{unravel,post_prl_long}.
 In this
work, we consider only the self-part of the total overlap
function (i.e. neglecting the $i \neq j$ terms in the double
summation). Earlier it has been shown to be a good approximation to the full
 overlap function. So,
the overlap function can be well approximated by its self part, 
and written as,  
\begin{eqnarray}
 \langle q(t) \rangle \approx \left \langle \sum_{i=1}^{N} \delta({\bf{r}}_i(t_0)-{\bf{r}}_i(t+t_0)) \right \rangle.
\end{eqnarray}

Again, the $\delta$ function is approximated by a Heaviside step function $\Theta(x)$ which defines the
condition of “overlap” between two particle positions
separated by a time interval t:
\begin{eqnarray}
 \langle q(t) \rangle \approx \left \langle \sum_{i=1}^{N} \Theta (\mid{\bf{r}}_i(t_0)-{\bf{r}}_i(t+t_0)\mid) \right \rangle \nonumber\\
\Theta(x) = 1, x \leq {\text{a implying “overlap”}} \nonumber\\
=0, \text{otherwise}.
\end{eqnarray}

The time dependent overlap function thus depends on
the choice of the cut-off parameter a, which we choose
to be 0.3. This parameter is chosen such that particle positions separated due to small amplitude vibrational motion are treated as the same, or that $a^2$ is comparable to
the value of the MSD in the plateau between the ballistic
and diffusive regimes.

Relaxation times obtained from the decay of the self intermediate scattering function $F_s (k, t)$ using the 
definition $F_s (k, t = \tau_\alpha)$ = $1/e$ at a fixed $k\simeq 2\pi /r_{max}$,  where $r_{max}$ is the position of the first peak of the radial
distribution function of A type particles. The self
intermediate scattering function is calculated from the
simulated trajectory as
\begin{equation}
 F_s(k,t)=\frac{1}{N}\left \langle \sum_{i=1}^{N} \exp(-i{\bf{k}}.({\bf{r}}_i(t)-{\bf{r}}_i(0))) \right \rangle.
\end{equation}

For NTW model, the cut off parameter for overlap function is taken as $a=0.2$ , as the decay time is similar to the decay observed from
 $F_s (k, t)$ at $k=8.2$\cite{szamel-ntw}.

Since relaxation times from $q(t)$ and $F_s (k, t)$
behave very similarly, we have used the time scale
obtained from q(t).

We find that the  diffusivities (D), obtained from the mean squared
displacement (MSD) of the particles, follow a similar trend like inverse of relaxation time. However, we have used 
the relaxation time data in this present work.

 \subsection{Excess Entropy}

 The thermodynamic excess entropy, $S_{ex}$, is defined as the difference between
the total entropy $(S_{total})$ and the ideal gas entropy $(S_{id})$ at
the same temperature $(T)$ and density $(\rho$) for all the model
glass formers\cite{murari_charu_2010_jcp,Murari_charu_jcp}.  The entropy was initially evaluated at a high temperature ($T_r$)
and low reduced density ($\rho_r$)
where the system can be assumed to behave as an ideal gas.
Entropies at any other state points, relative to this reference ideal
state point, can be calculated using a combination of isochoric
and isothermal paths, ensuring that no phase transitions occur
along the path. 
The entropy of the liquid has been calculated via thermodynamic integration
starting from the ideal gas binary mixture reference point, along the $T = T_r$ isotherm,
up to the studied $\rho = \rho_{t}$ density.

The entropy change along an isothermal path
is given by \cite{Murari_charu_jcp},
\begin{widetext}
\begin{equation}
 S_{ex}(T_r, \rho_{t})=S(T_r,\rho_{t})- S(T_r,\rho_{r})=\frac{U(T_r,\rho_r)- U(T_r,\rho_t)}{T_r}-\int_{\rho_r}^{\rho_t}\frac{P(\rho)}{T_r}\frac{N}{\rho^2}d\rho,
\end{equation}
and along the isochoric path it is given by \cite{Murari_charu_jcp},

\begin{equation}
 S_{ex}(T_t, \rho_{t})=S(T_t,\rho_t)- S(T_r,\rho_t)=\int_{T_r}^{T_t}\frac{1}{T} \left(\frac{\partial U}{\partial T}\right)_\rho dT.
\end{equation}

\end{widetext}

To calculate the integral in the above equation, we use a standard technique used for supercooled liquid \cite{Srikanth_PRL}. We fit the temperature dependence of the potential energy using the 
functional form, $U=a+bT^c$, where $a,b,c$ are the fitting parameters. The low temperature trajectories have been averaged over $500\tau$ (where $\tau$
is the relaxation time defined in Sec 3A).

\section{Results}

\begin{figure}[h]
\centering
 \includegraphics[width=0.45\textwidth]{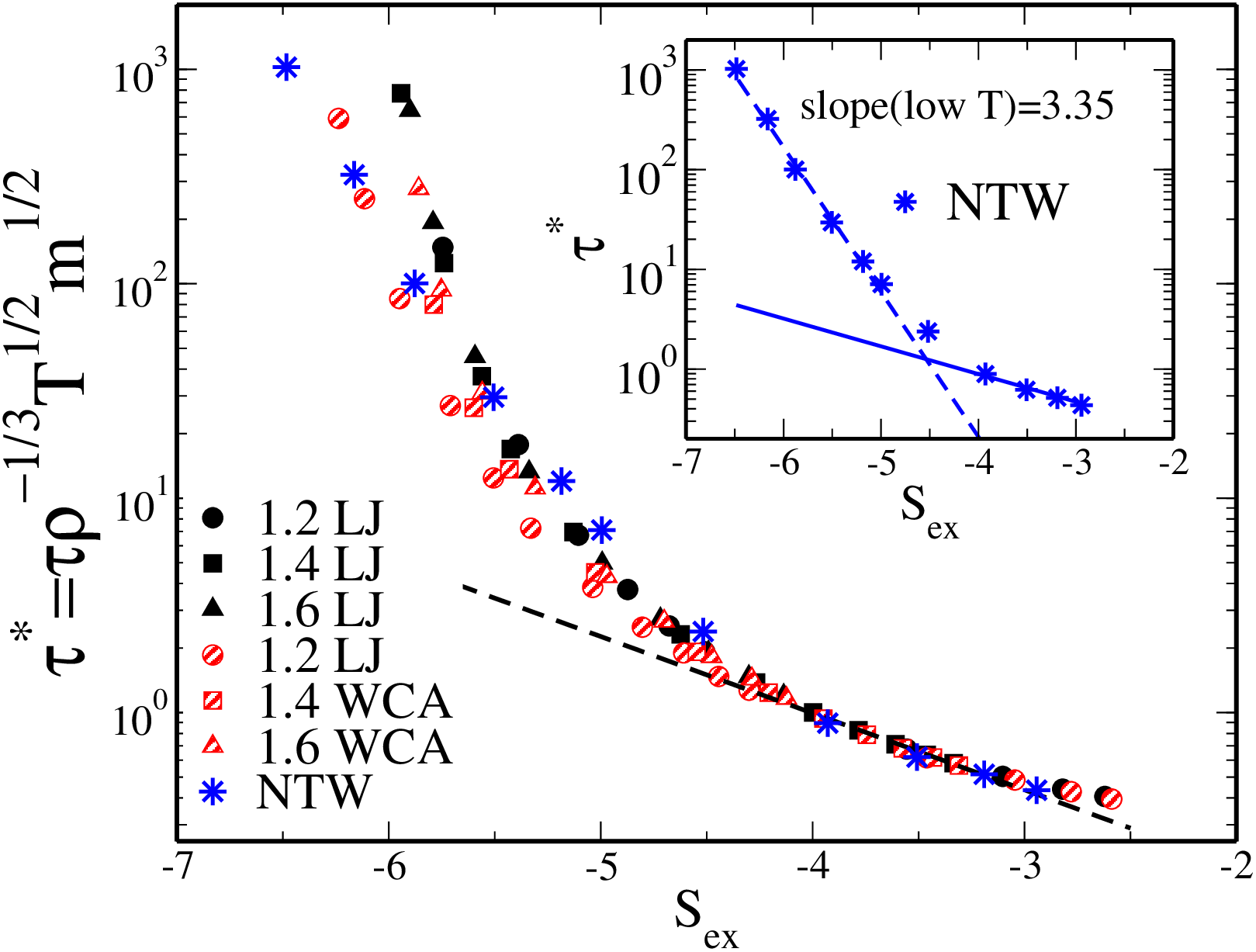}
\caption{The scaled relaxation time vs. excess entropy ($S_{ex}$). In the high temperature regime, the Rosenfeld  scaling law follow a universal behavior with the exponent 0.8.
The $\ln \tau^*$ is shifted by $-0.2$ for NTW model to obtain the master plot. Note that, the slope remains unaltered with this shifting. The slope of the black dashed line is close to 0.8.
(inset) For the NTW model the scaled relaxation time vs. excess entropy ($S_{ex}$). The low temperature
regime is fitted to another straight line with different slope which is equal to $3.35$.}
\label{scale_tau_vs_Sex}
\end{figure}

\begin{figure}[h]
\centering
 \includegraphics[width=0.45\textwidth]{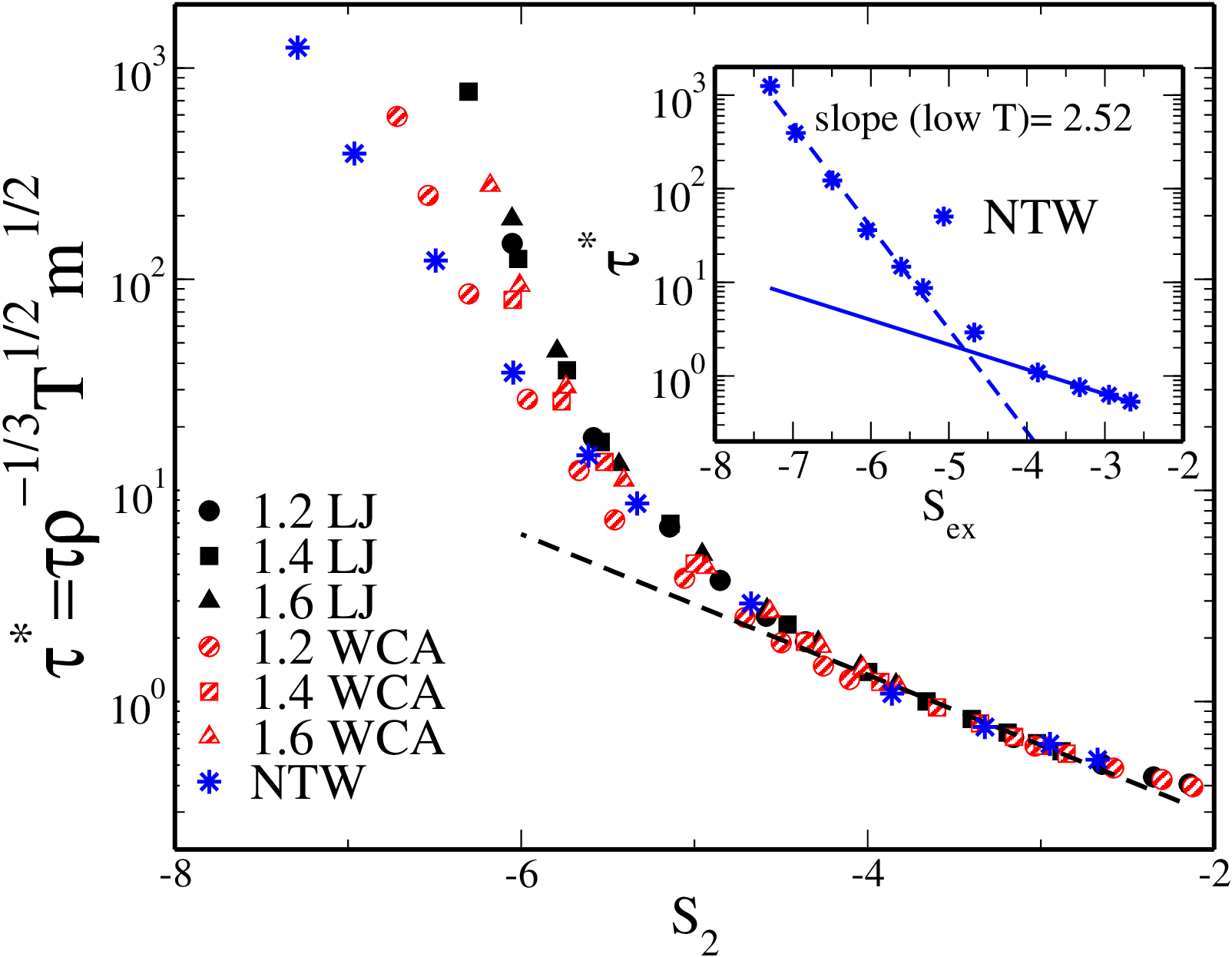}
\caption{The  scaled relaxation time vs. pair entropy ($S_{2}$). In the high temperature regime, the Dzugutov  scaling law follow a universal behavior with the exponent 0.8.
 The slope of the black dashed line is close to 0.8. (inset) For the NTW model the scaled relaxation time vs. pair entropy ($S_{2}$). The low temperature
regime is also fitted to another straight line with different slope which is  equal to $2.52$.}
\label{scale_tau_vs_S2}
\end{figure}

In case of simple liquids at the high temperatures, the dimensionless scaled relaxation time ($\tau^*$) follows the Rosenfeld relation which is  given by,
\begin{equation}
 \tau^* =C \exp (-K S_{ex}),
\end{equation}

\noindent where, $C$, $K$ are the fitting parameters and $S_{ex}$ is the excess entropy. 
The scaled relaxation time $\tau^*$ can be written as $\tau^*=\tau \rho^{-1/3}(k_B T/m)^{-1/2}$, where $m$ is the mass of the particle. Note that, for NTW model we 
have used the mass as $m=\sum_i x_i m_i$, where $x_i$ is the mole fraction of the i-th particle.
 We plot the Rosenfeld behavior both for LJ and WCA systems at different
densities and for the NTW model at $\rho=1.655$ (Fig \ref{scale_tau_vs_Sex}). We find that the relationship is valid at high temperature regime. 
Interestingly all the systems show a master plot at high temperature and
the Rosenfeld exponent ($K$) is close to $0.8$. 
Note that earlier studies have reported that the Rosenfeld scaling is valid at high temperature both for LJ and WCA systems with the same value of the exponent $K$
\cite{Murari_charu_jcp}.
However the density range studied here is much higher, thus, predicting that the Rosenfeld relationship is valid over a wider range of densities.
Our study also shows that although at low temperatures the WCA system does not follow any density-temperature scaling
\cite{tarjus_epje,Pedersen-prl,unravel},
at high temperatures it shows a master plot. The NTW model on the other hand, although is a model of a simple liquid, is known to show characteristics of ionic
melts, to be specific the $SiO_2$ system \cite{coslovich_pastore_jpcm_ntw}. However the value of the Rosenfeld scaling exponent ($K$) for NTW model is similar to that of the simple liquids and
different from that reported for $SiO_2$ \cite{agarwal2009relationship}. 

  Our next observation is that in the low temperature regime although the LJ and WCA systems show a breakdown of the scaling law, the NTW model follows the scaling law with 
higher
value of the  exponent (inset of Fig \ref{scale_tau_vs_Sex}). This behavior of the NTW model is 
similar to that reported for the ionic melts \cite{agarwal2009relationship}.
 Thus, our study reveals that although the NTW model follows the Rosenfeld behaviour up 
to a much lower temperature like other
network forming liquids, its Rosenfeld exponent is similar to that found for simple liquids ($K\sim0.8$). Therefore, 
the NTW model appears to have mixed characteristics of network
forming liquids and simple liquids.

 As mentioned in the introduction, the pair entropy $S_2$  provides $80\%-90\%$ contribution to
 the excess entropy \cite {Baranyai-cp, murari_charu_2010_jcp,unravel} and thus, the Rosenfeld relationship can be written in terms of $S_2$ which is given by,
\begin{equation}
 \tau^* =C_1 \exp (-K_1 S_{2}),
\end{equation}
where, $C_1$, $K_1$ are the fitting parameters and $S_{2}$ is the pair entropy and this scaling behavior is known as Dzugutov scaling law. In Fig \ref{scale_tau_vs_S2}
we plot the Dzugutov scaling behavior and similar to the Rosenfeld relationship we find it to be valid at high temperature regime.  
The value of the exponent $K_1 \sim 0.8$, this indicates that both the Rosenfeld
and Dzugutov scaling follows an universal exponent.  We also find that  similar to the Rosenfeld relation,
the Dzugutov scaling behavior is valid for NTW model upto a much lower temperature (inset of Fig \ref{scale_tau_vs_S2}) with a higher value of exponent.

\begin{figure}[h]
\centering
\subfigure{
 \includegraphics[width=0.4\textwidth]{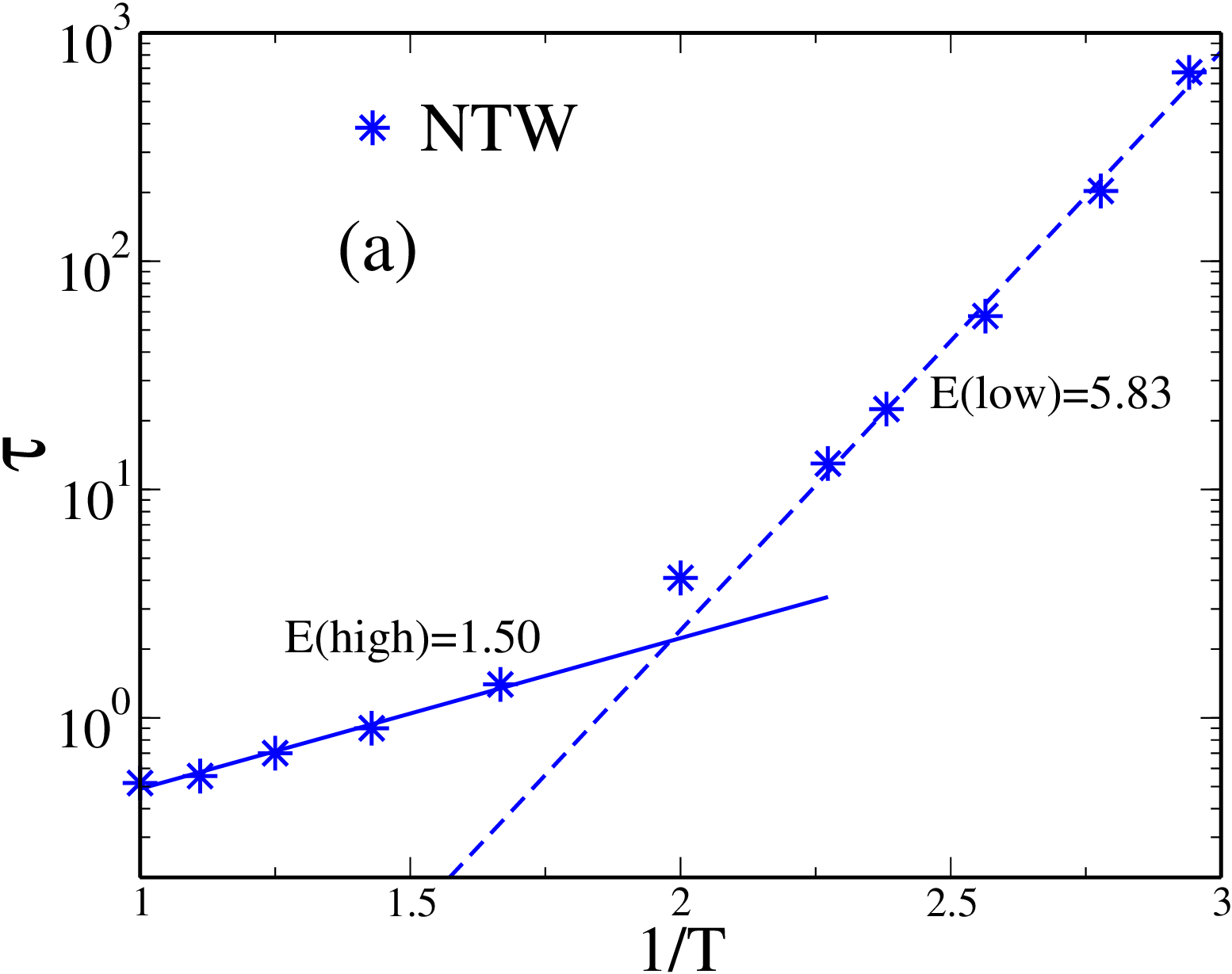}}
\subfigure{
 \includegraphics[width=0.4\textwidth]{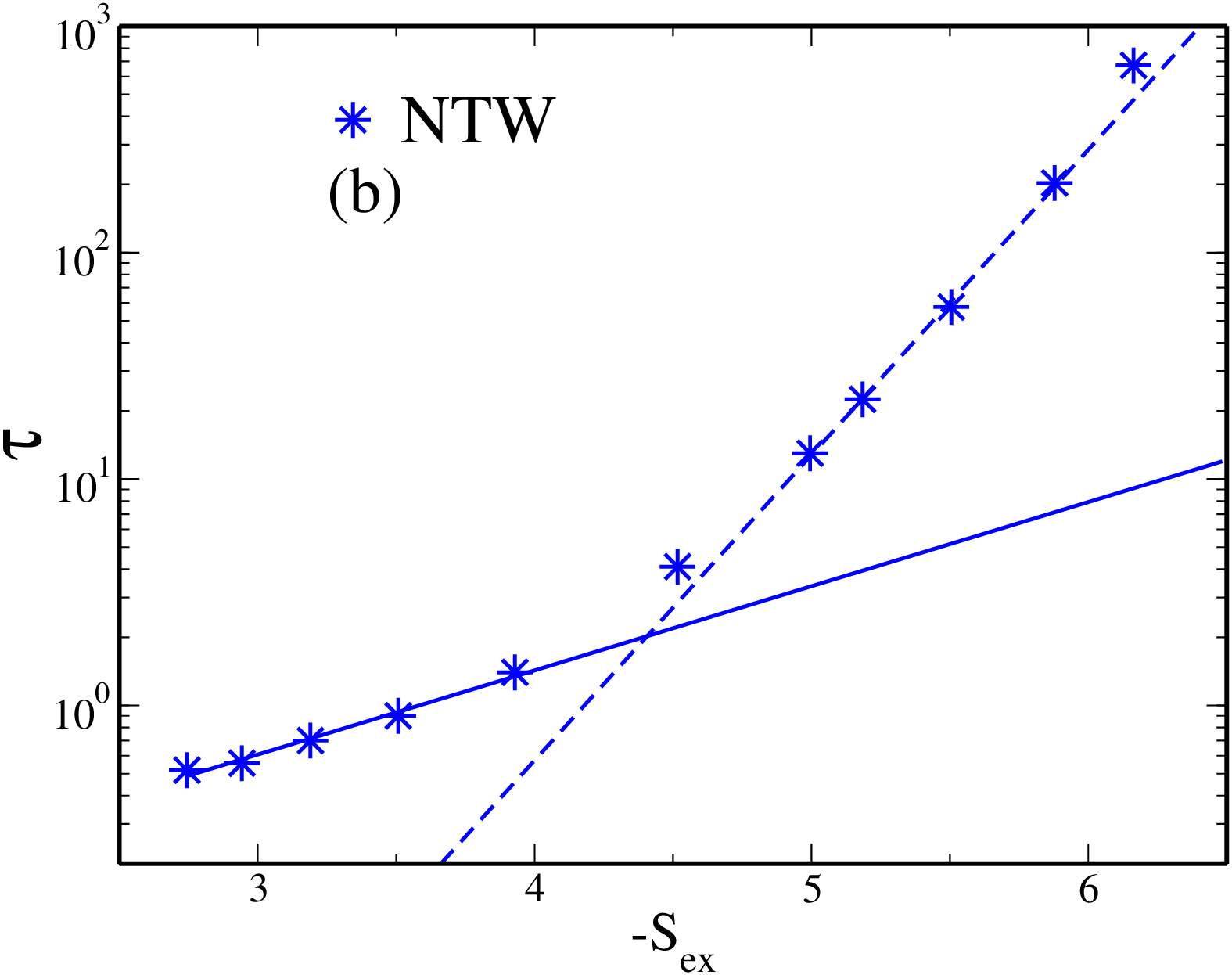}}
\subfigure{
\includegraphics[width=0.4\textwidth]{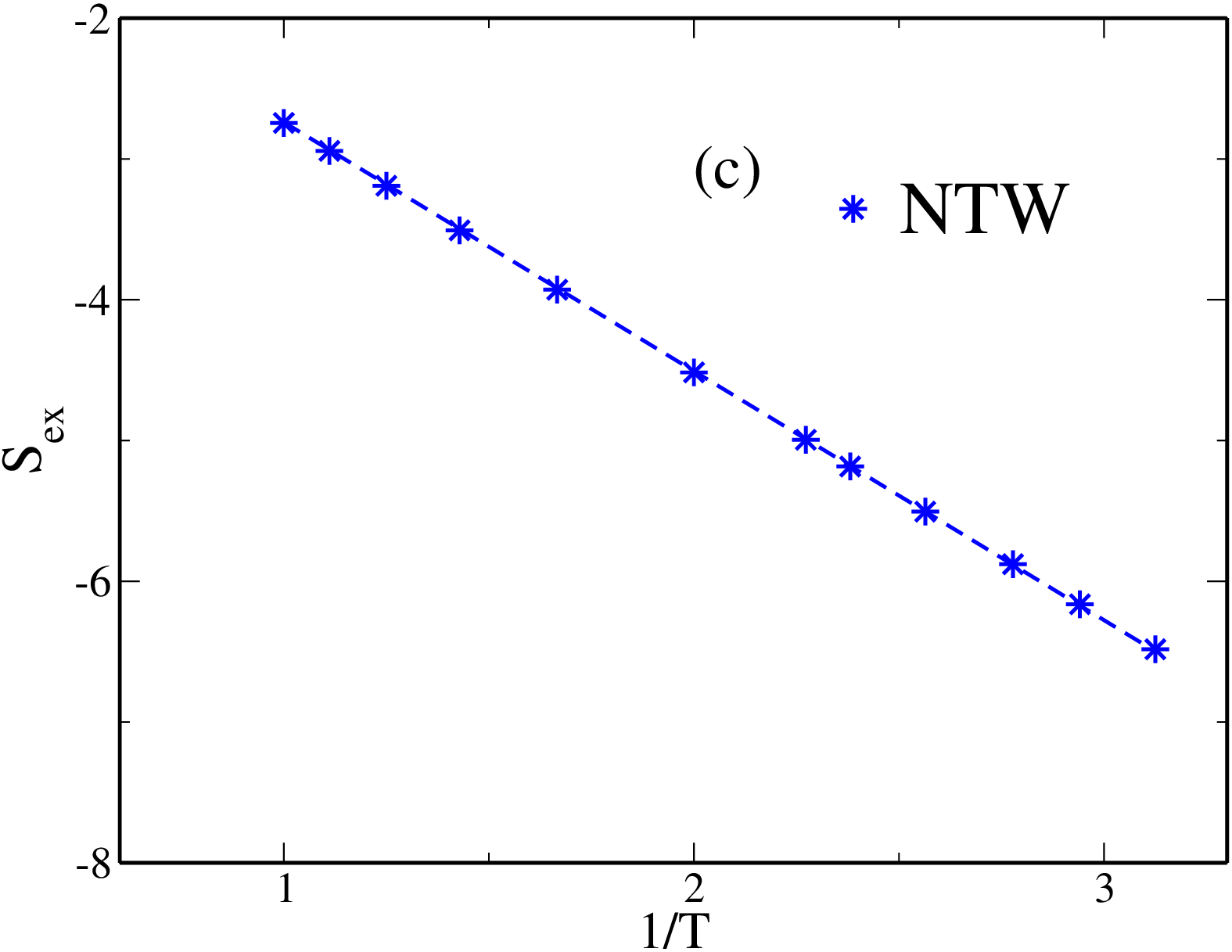}}
\caption{ For the NTW model (a) the temperature dependence of the relaxation time. Both the high and low temperature regimes follow Arrhenius relationship and can be fitted to 
two independent straight lines. (b) The validity of the Rosenfeld relationship. The break in both the plots (a and b) occur at the same value of the $\tau$. (c) $S_{ex}$ vs $1/T$ shows a linear
behavior over the whole temperature regime.  }
\label{NTW-tau-rosen-linear}
\end{figure}

Next we explore the origin of validity of the Rosenfeld  scaling for the network forming liquid even in the lower temperature regime.
For any system 
at high temperatures, the temperature dependence of the relaxation time can be expressed in terms of Arrhenius law which is given by,
\begin{equation}
 \tau = \tau_0 \exp (E/T),
\label{arh}
\end{equation}
where, E is the activation energy. Note that the ionic melts and NTW model are strong liquids which follow Arrhenius behaviour even in the lower temperature
regime.
In Fig \ref{NTW-tau-rosen-linear}a we plot the relaxation time  for NTW model and find that both the high temperature and low temperature regimes can be fitted to two 
independent straight lines with the activation energies $E (high)=1.50$  and $E (low)=5.83$, respectively.

The Rosenfeld  scaling law can also be written in terms of unscaled relaxation time \cite{unravel} given by,
\begin{eqnarray}
 \tau = C^\prime \exp (-K^\prime S_{ex}). 
\label{rosenfeld-unscaled}
\end{eqnarray}
We plot the $\tau$ vs $-S_{ex}$  in Fig \ref{NTW-tau-rosen-linear}b and find that the Rosenfeld  behavior is valid up to a low temperature regime with different values of the
exponent which are given in Table \ref{parameters}. Note that the break in the Rosenfeld behaviour happens exactly at the same position where the break in the Arrhenius
behavior is observed. Thus, these two plots show qualitatively similar behavior.   

\begin{table}[h]
\caption{ The fitting parameters of the Rosenfeld scaling,$\tau = C^\prime \exp (-K^\prime S_{ex})$.}
 \begin{tabular}{|c|c|c|}
 \hline
    NTW & $\ln C^\prime$  &$K^\prime$    \\ \hline
    high T&3.07 & 0.86 \\ \hline
    low T & 12.98 &3.10  \\ \hline
\end{tabular}
\label{parameters}
\end{table}

Now to establish a quantitative connection between the Arrhenius and the Rosenfeld relations, we equate Eq \ref{arh} and Eq \ref{rosenfeld-unscaled} and can write it as, 
\begin{equation}
 \frac{E}{T}\sim -K^\prime S_{ex}.
\label{equality}
\end{equation}
The above equation suggests that for strong liquids where E is independent of temperature if $S_{ex} \propto 1/T$, 
the activation energy (E)  is proportional to the Rosenfeld exponent ($K^\prime$).

Next we show, that for NTW model the $S_{ex}$ vs. $1/T$ plot indeed shows a linear behavior (Fig \ref{NTW-tau-rosen-linear}c) for the whole temperature range. 
Thus, from Eq \ref{equality} we expect that
\begin{equation}
 \frac{E (low)}{E (high)} \sim \frac{K^\prime(low)}{K^\prime(high)}.
\end{equation}
We calculate the ratio of the activation energies and find that $\frac{E (low)}{E (high)}=3.89$ which is close to the value of
 $\frac{K^\prime(low)}{K^\prime(high)}=3.60$.

Note that for most of the systems both  the Rosenfeld relationship and the Arrhenius behaviour are  known to be valid only in the high temperature regime.
For the fragile systems, in the low temperature regime, the activation energy becomes temperature dependent and there is a breakdown of the Arrhenius behaviour.
However, strong liquids like the NTW model follows an Arrhenius temperature dependence even in the low temperature regime (Fig \ref{NTW-tau-rosen-linear}a) 
and also follows the Rosenfeld scaling (Fig \ref{NTW-tau-rosen-linear}b). 
 Thus we may conclude that the breakdown of the Rosenfeld relationship is connected to the 
temperature dependence of the activation energy, E.

\begin{figure}[h]
\centering
 \includegraphics[width=0.45\textwidth]{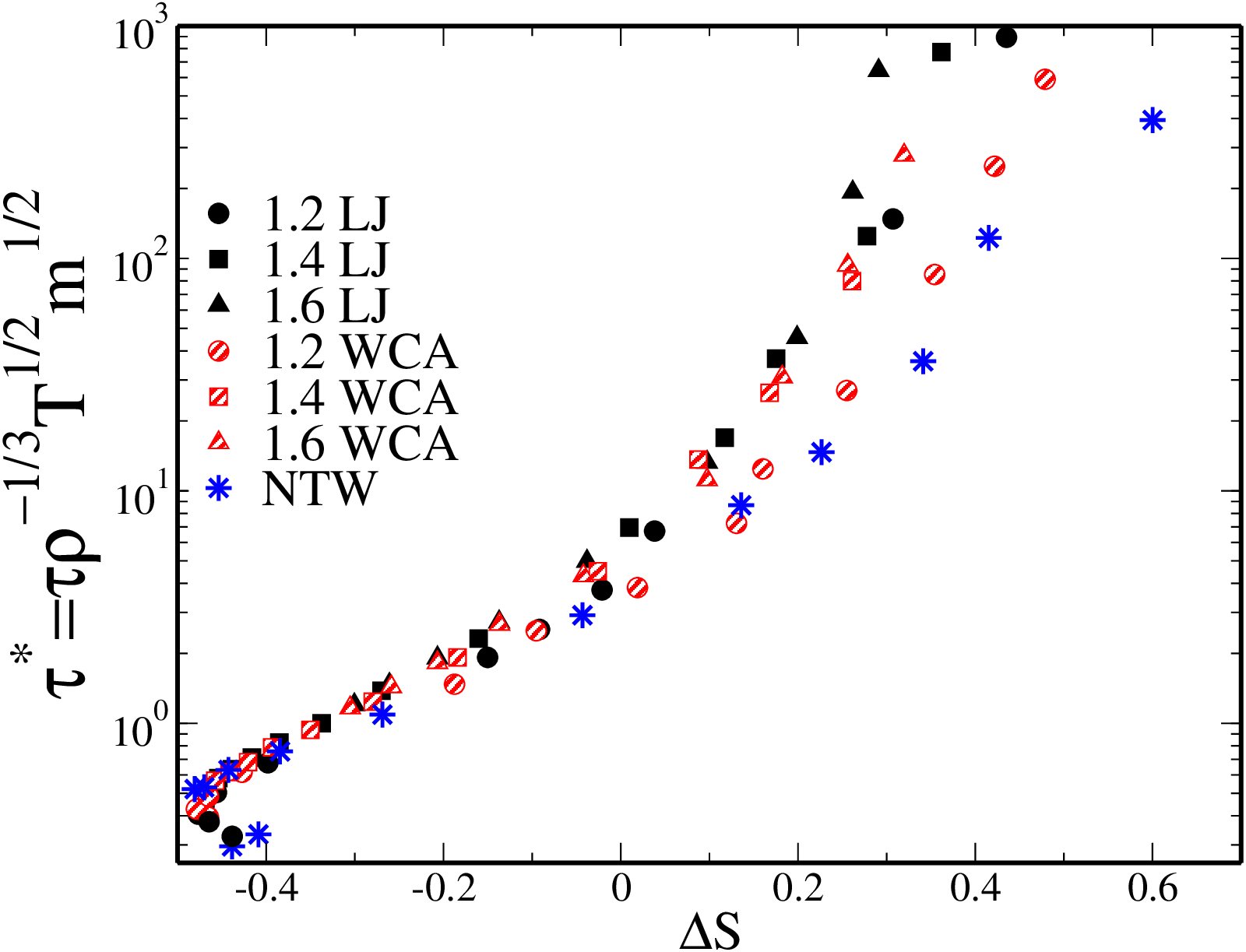}
\caption{The scaled relaxation time vs. RMPE ($\Delta S$). To start with the same high temperature value of $\Delta S$, for NTW model the x-axis is
shifted by $-0.2$ .}
\label{scale_tau_vs_RMPE}
\end{figure}

\begin{figure}[h]
\centering
 \includegraphics[width=0.45\textwidth]{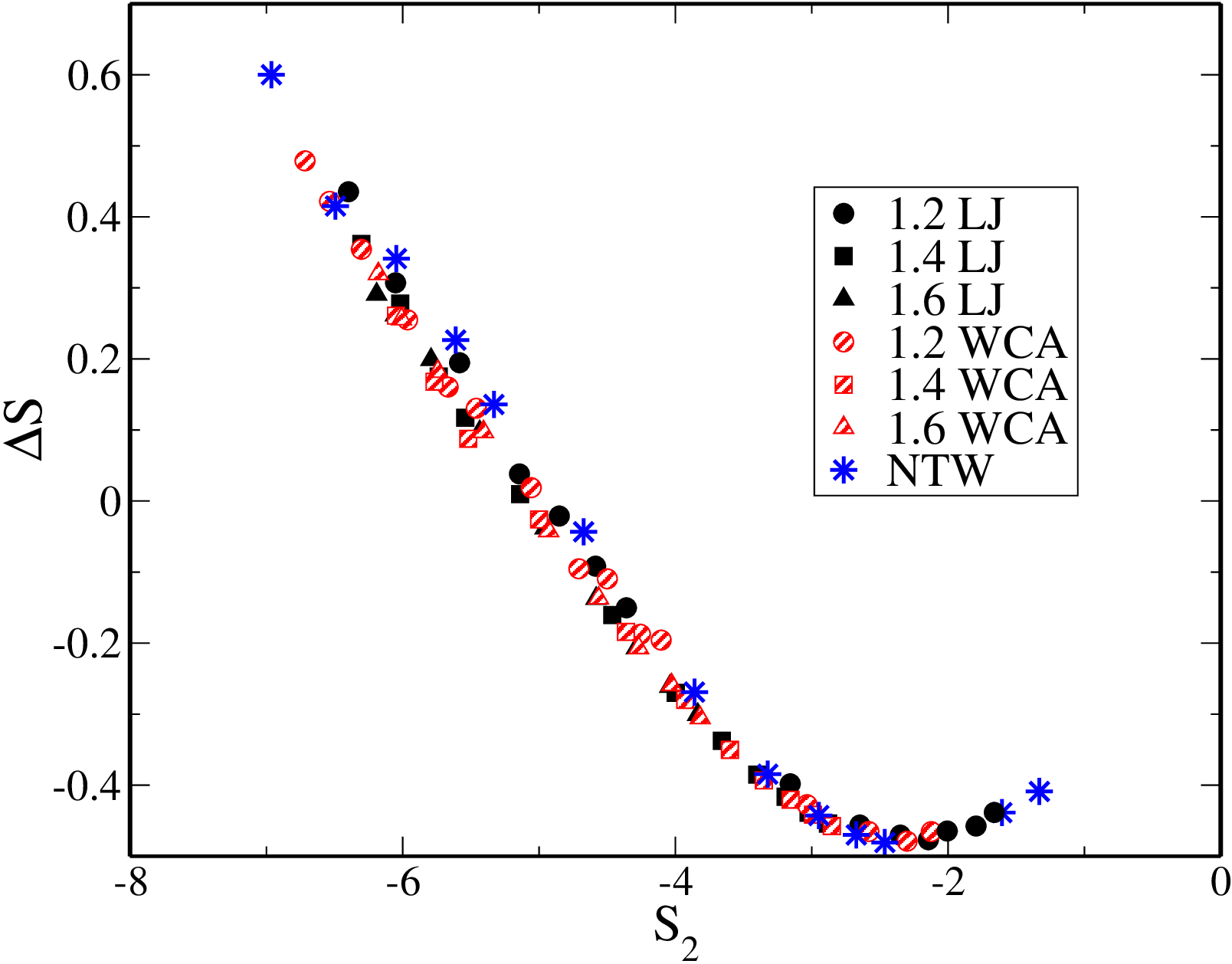}
\caption{Correlation of RMPE ($\Delta S$) with pair entropy($S_2$). For NTW model the $\Delta S$ (y-axis) is
shifted by $-0.2$ .}
\label{S2_vs_RMPE}
\end{figure}

Next we explore the role of residual multi particle entropy (RMPE)  in the dynamics.  In Fig \ref{scale_tau_vs_RMPE} we plot
the scaled relaxation time with RMPE, $\Delta S$. Similar to that found earlier \cite{Murari_charu_jcp}, for all the systems,
 there is initially a positive correlation between the relaxation time and $\Delta S$ and as the temperature is lowered the correlation becomes negative.
This role reversal of RMPE has been observed earlier where we 
have reported that the small positive value of RMPE speeds up the relaxation time \cite{bssb}.
Although at moderately high temperatures, the data show a master plot, at low temperature the dependence of relaxation time on  RMPE becomes system dependent.
Similar observation is made in Fig \ref{scale_tau_vs_S2} where $\tau^*$ vs. $S_2$ also shows a deviation at low temperature.

Note that these systems cover a wide range of fragility where the LJ system and WCA at high density are the most fragile systems and NTW model is the least fragile one (strongest).
Thus, the $S_2$ and $\Delta S$ dependence of the scaled relaxation time $\tau^*$ show a variation with fragility. It appears that for the strong liquid, there is a weaker
dependence of $\tau^*$ on $S_2$ and $\Delta S$.

  To understand this, in Fig \ref{S2_vs_RMPE} we plot $\Delta S$ with pair entropy for all the systems and find a data collapse. The plot shows that
at high temperatures, there is  a positive correlation of $S_2$ and RMPE up to a point where $S_2=-2.5K_B$ \cite{Murari_charu_jcp}.
 Below $S_2=-2.5 K_B$, $S_2$  decreases with increasing $\Delta S$ and the rate of the negative correlation is
independent of fragility.
Since the $S_2$ and the $\Delta S$ act in opposite directions, their combined effect which is seen in the $S_{ex}$ and thus in the relaxation time is much weaker.
This is precisely the reason behind the spread observed in Fig \ref{scale_tau_vs_S2} and Fig \ref{scale_tau_vs_RMPE}.
Note that $\tau^*$
vs. $S_{ex}$ plot shows a less spread with fragility (Fig \ref{scale_tau_vs_Sex}).

\section{Conclusion}
In this present work we find that both the Rosenfeld and Dzugutov scaling laws are valid at the high temperature regime for LJ and WCA systems ($\rho=1.2-1.6$) and 
for the NTW model ($\rho=1.655$). Interestingly, for all the systems the exponents of both the 
scaling laws are universal at high temperatures\cite{Murari_charu_jcp} and different from the value predicted for the ionic melts \cite{agarwal2009relationship}.
For the LJ and the WCA systems both the scaling laws break down in the low temperature regime.
However, we show that like other network forming liquids (ionic melts)\cite{agarwal2009relationship}, the scaling laws for the NTW model
are valid  even in the lower temperature regime. Thus, although in terms of validity of these
semi-empirical relationships, the NTW model behaves like other network forming liquids, the value of the exponent appears to be similar to that found for the other simple 
liquids. Our study establishes that the Rosenfeld and the Arrhenius relations are correlated both qualitatively and quantitatively. 
Thus, for the NTW model, the validity of the Rosenfeld relationship  at lower temperatures can be connected to it being a strong liquid following Arrhenius
behaviour in this regime. We also study the independent role of pair entropy and residual multiparticle entropy on the dynamics for the systems 
with different fragilities. Our study reveals that for stronger liquids the dynamics has a weaker dependence on the $S_2$ and $\Delta S$. This weaker dependence
can be traced back to the fact that the effect of $S_2$ and $\Delta S$ on the dynamics act in the opposite directions with a similar rate.

\section{Acknowledgements}
This work is dedicated to the memory of Professor Charusita Chakravarty. 
A. Banerjee and M. K. Nandi contributed equally to this work.
SMB thanks Charusita Chakravarty and Srikanth Sastry for discussions.
Authors thank Department of Science and Technology (DST-SERB), India and CSIR-Multi-Scale Simulation and Modeling project for financial supports.
AB thanks CSIR and MKN thanks UGC for fellowship.

\end{document}